\documentclass{nature}

\usepackage{amssymb}
\usepackage{graphicx}
\usepackage{color}
\usepackage[normalem]{ulem} 
\usepackage{soul}

\usepackage{subfig}
\usepackage[font=footnotesize]{caption}

\usepackage{floatrow}
\DeclareFloatFont{tiny}{\scriptsize}
\title{Acceleration of Petaelectronvolt protons in the Galactic Centre}

\newcounter{firstbib}

\begin{document}

\maketitle

\author{H.E.S.S. Collaboration:
A.~Abramowski$^{1}$,
F.~Aharonian$^{2,3,4}$,
F.~Ait Benkhali$^{2}$,
A.G.~Akhperjanian$^{5,4}$,
E.O.~Ang\"uner$^{6}$,
M.~Backes$^{7}$,
A.~Balzer$^{8}$,
Y.~Becherini$^{9}$,
J.~Becker Tjus$^{10}$
% ...
}

\begin{affiliations}
\small
\item Universit\"at Hamburg, Institut f\"ur Experimentalphysik, Luruper Chaussee 149, D 22761 Hamburg, Germany 
\item Max-Planck-Institut f\"ur Kernphysik, P.O. Box 103980, D 69029 Heidelberg, Germany 
\item Dublin Institute for Advanced Studies, 31 Fitzwilliam Place, Dublin 2, Ireland 
\item National Academy of Sciences of the Republic of Armenia,  Marshall Baghramian Avenue, 24, 0019 Yerevan, Republic of Armenia  
\item Yerevan Physics Institute, 2 Alikhanian Brothers St., 375036 Yerevan, Armenia 
\item Institut f\"ur Physik, Humboldt-Universit\"at zu Berlin, Newtonstr. 15, D 12489 Berlin, Germany 
\item University of Namibia, Department of Physics, Private Bag 13301, Windhoek, Namibia 
\item GRAPPA, Anton Pannekoek Institute for Astronomy, University of Amsterdam,  Science Park 904, 1098 XH Amsterdam, The Netherlands 
\item Department of Physics and Electrical Engineering, Linnaeus University,  351 95 V\"axj\"o, Sweden 
\item Institut f\"ur Theoretische Physik, Lehrstuhl IV: Weltraum und Astrophysik, Ruhr-Universit\"at Bochum, D 44780 Bochum, Germany 
\item GRAPPA, Anton Pannekoek Institute for Astronomy and Institute of High-Energy Physics, University of Amsterdam,  Science Park 904, 1098 XH Amsterdam, The Netherlands 
\item Institut f\"ur Astro- und Teilchenphysik, Leopold-Franzens-Universit\"at Innsbruck, A-6020 Innsbruck, Austria 
\item School of Chemistry \& Physics, University of Adelaide, Adelaide 5005, Australia 
\item Centre for Space Research, North-West University, Potchefstroom 2520, South Africa 
\item LUTH, Observatoire de Paris, CNRS, Universit\'e Paris Diderot, 5 Place Jules Janssen, 92190 Meudon, France 
\item LPNHE, Universit\'e Pierre et Marie Curie Paris 6, Universit\'e Denis Diderot Paris 7, CNRS/IN2P3, 4 Place Jussieu, F-75252, Paris Cedex 5, France 
\item Laboratoire Univers et Particules de Montpellier, Universit\'e Montpellier 2, CNRS/IN2P3,  CC 72, Place Eug\`ene Bataillon, F-34095 Montpellier Cedex 5, France 
\item DSM/Irfu, CEA Saclay, F-91191 Gif-Sur-Yvette Cedex, France 
\item Astronomical Observatory, The University of Warsaw, Al. Ujazdowskie 4, 00-478 Warsaw, Poland 
\item Aix Marseille Universi\'e, CNRS/IN2P3, CPPM UMR 7346,  13288 Marseille, France 
\item Instytut Fizyki J\c{a}drowej PAN, ul. Radzikowskiego 152, 31-342 Krak{\'o}w, Poland 
\item Funded by EU FP7 Marie Curie, grant agreement No. PIEF-GA-2012-332350,  
\item School of Physics, University of the Witwatersrand, 1 Jan Smuts Avenue, Braamfontein, Johannesburg, 2050 South Africa 
\item Landessternwarte, Universit\"at Heidelberg, K\"onigstuhl, D 69117 Heidelberg, Germany 
\item Oskar Klein Centre, Department of Physics, Stockholm University, Albanova University Center, SE-10691 Stockholm, Sweden 
\item Wallenberg Academy Fellow,  
\item Institut f\"ur Astronomie und Astrophysik, Universit\"at T\"ubingen, Sand 1, D 72076 T\"ubingen, Germany 
\item Laboratoire Leprince-Ringuet, Ecole Polytechnique, CNRS/IN2P3, F-91128 Palaiseau, France 
\item APC, AstroParticule et Cosmologie, Universit\'{e} Paris Diderot, CNRS/IN2P3, CEA/Irfu, Observatoire de Paris, Sorbonne Paris Cit\'{e}, 10, rue Alice Domon et L\'{e}onie Duquet, 75205 Paris Cedex 13, France 
\item Univ. Grenoble Alpes, IPAG,  F-38000 Grenoble, France \\ CNRS, IPAG, F-38000 Grenoble, France 
\item Department of Physics and Astronomy, The University of Leicester, University Road, Leicester, LE1 7RH, United Kingdom 
\item Nicolaus Copernicus Astronomical Center, ul. Bartycka 18, 00-716 Warsaw, Poland 
\item Institut f\"ur Physik und Astronomie, Universit\"at Potsdam,  Karl-Liebknecht-Strasse 24/25, D 14476 Potsdam, Germany 
\item Laboratoire d'Annecy-le-Vieux de Physique des Particules, Universit\'{e} Savoie Mont-Blanc, CNRS/IN2P3, F-74941 Annecy-le-Vieux, France 
\item DESY, D-15738 Zeuthen, Germany 
\item Obserwatorium Astronomiczne, Uniwersytet Jagiello{\'n}ski, ul. Orla 171, 30-244 Krak{\'o}w, Poland 
\item Universit\'e Bordeaux, CNRS/IN2P3, Centre d'\'Etudes Nucl\'eaires de Bordeaux Gradignan, 33175 Gradignan, France 
\item Universit\"at Erlangen-N\"urnberg, Physikalisches Institut, Erwin-Rommel-Str. 1, D 91058 Erlangen, Germany 
\item Centre for Astronomy, Faculty of Physics, Astronomy and Informatics, Nicolaus Copernicus University,  Grudziadzka 5, 87-100 Torun, Poland 
\item Department of Physics, University of the Free State,  PO Box 339, Bloemfontein 9300, South Africa 
\item Heisenberg Fellow (DFG), ITA Universit\"at Heidelberg, Germany,  
\item GRAPPA, Institute of High-Energy Physics, University of Amsterdam,  Science Park 904, 1098 XH Amsterdam, The Netherlands
\end{affiliations}

\hfill

\begin{abstract}
Galactic cosmic rays reach energies of at least a few Peta-electronvolts (1 PeV =$10^\mathbf{15}$ electron volts)\cite{1990acr..book.....B}. This implies our Galaxy contains PeV accelerators ({\it PeVatrons}), but all proposed models of Galactic cosmic-ray accelerators encounter non-trivial difficulties at exactly these energies\cite{2001RPPh...64..429M}. Tens of Galactic accelerators capable of accelerating particle to tens of TeV (1 TeV =$10^\mathbf{12}$ electron volts) energies were inferred from recent gamma-ray observations\cite{2013APh....43...19H}. None of the currently known accelerators, however, not even the handful of shell-type supernova remnants commonly believed to supply most Galactic cosmic rays, have shown the characteristic tracers of PeV particles: power-law spectra of gamma rays extending without a cutoff or a spectral break to tens of TeV\cite{2013APh....43...71A}. Here we report deep gamma-ray observations with arcminute angular resolution of the Galactic Centre regions, which show the expected tracer of the presence of PeV particles within the central 10~parsec of the Galaxy.  We argue that the supermassive black hole Sagittarius A* is linked to this {\it PeVatron}.  Sagittarius A* went through active phases in the past, as demonstrated by X-ray outbursts\cite{Clavel:2013bfa} and an outflow from the Galactic Center\cite{Su2010ApJ}. Although its current rate of particle acceleration is not sufficient to provide a substantial contribution to Galactic cosmic rays, Sagittarius A* could have plausibly been more active over the last $\gtrsim 10^{6-7}$~years, and therefore should be considered as a viable alternative to supernova remnants as a source of PeV Galactic cosmic rays.
\end{abstract}

The large photon statistics accumulated over the last 10 years of observations with the High Energy Stereoscopic System (H.E.S.S.),  together with improvements in the methods of data analysis,  allow for a deep study of the properties of  the diffuse very-high-energy (VHE; more than 100 GeV)  emission of the central molecular zone. This region surrounding the Galactic Centre contains predominantly molecular gas and extends (in projection) out to r$\sim$250 pc at positive galactic longitudes and r$\sim$150 pc at negative longitudes. The map of the central molecular zone as seen in VHE $\gamma$-rays  (Fig.~1) shows a  strong (although not linear; see below) correlation between the brightness distribution of VHE $\gamma$-rays and the locations of massive gas-rich complexes. This points  towards a hadronic origin of the diffuse emission\cite{Aharonian:2006au}, where the 
$\gamma$-rays result from the interactions of relativistic protons with the ambient  gas. The second important mechanism of production of VHE $\gamma$-rays is the inverse Compton scattering of  electrons. However, the severe  radiative losses suffered by multi-TeV electrons in the Galactic Centre region prevent them from propagating over scales comparable to the size of the central molecular zone, thus disfavouring a leptonic origin of the $\gamma$-rays  (see discussion in Methods and Extended Data Figures~1 and~2).

The location and the particle injection rate history of the cosmic-ray accelerator(s), responsible for the relativistic protons, determine the spatial distribution of these cosmic rays which, together with the gas distribution, shape  the  morphology of the central molecular zone seen in VHE  $\gamma$-rays.  Fig.~2 shows the radial profile of the $E \geq 10$~TeV cosmic-rays energy density $w_{\rm CR}$ up to $r \sim 200$~pc (for a Galactic Centre distance of 8.5 kpc), determined from the $\gamma$-ray luminosity and the amount of target gas (see Extended Data Tables 1 and 2). This high energy density in the central molecular zone is found to be an order of magnitude larger than that of the ``sea'' of  cosmic rays that universally fills the Galaxy, while the energy density of low energy (GeV) cosmic rays in this region has a level comparable to it\cite{Yang}. This requires the presence of one or more accelerators of multi-TeV particles operating in the central molecular zone. 

If  the accelerator injects  particles  at a continuous rate, $\dot{Q}_p(E)$, the radial distribution of cosmic rays   in the central molecular zone,  in the case of diffusive propagation,  is described as  $w_{\rm CR}(E,r,t)=\frac{\dot{Q}_p(E)}{4 \pi D(E) r}$ erfc$(r/r_{\rm diff})$\cite{FA04}, where 
 $D(E)$ and $r_{\rm diff}$ are  the diffusion coefficient and radius, respectively. For timescales smaller than  the  proton-proton   interaction time in the hydrogen gas of density $n$, $t \leq t_{\rm pp} \simeq  5 \times 10^4 (n/10^3 \ \rm cm^{-3})^{-1} \ \rm yr$, the diffusion radius is $r_{\rm diff} \approx \sqrt{4 D(E) t}$. Thus, at distances $r < r_{\rm diff}$,  the  proton flux  should decrease as $\sim 1/r$  provided that the diffusion coefficient does not  significantly vary  throughout the  central molecular zone. The measurements  clearly support the  $w_{\rm CR}(r) \propto 1/r$  dependence over the entire central molecular zone region (Fig. 2) and disfavour a $w_{\rm CR}(r) \propto 1/r^2$  and a $w_{\rm CR}(r) \propto  constant$ profiles, as expected if cosmic rays are advected in a wind, and in the case of a single burst-like event of cosmic-ray injection, respectively. The $1/r$ profile of  the cosmic-ray density up to 200 pc  indicates a quasi-continuous injection of protons into the central molecular zone from a centrally located accelerator on a timescale $\Delta t$ exceeding the characteristic time of  diffusive escape of particles from the central molecular zone, {\textit i.e.} $\Delta t \geq t_{\rm diff} \approx R^2/6 D \approx  2 \times 10^3 (D/10^{30} \ \rm cm^2 s^{-1})^{-1}$~yr, where $D$ is normalised to  the characteristic value of  multi-TeV cosmic rays in the Galactic Disk\cite{strong}.  In this regime  the average injection rate of particles is found to be $\dot{Q}_p(\geq 10 \ \rm TeV) \approx 4 \times 10^{37} (D/10^{30} \ \rm cm^2 s^{-1})$ erg/s.  The diffusion  coefficient itself  depends on the power spectrum of the turbulent magnetic field, which is highly unknown in the  central molecular zone region. This   introduces  an uncertainty in  the estimates of  the  injection power of relativistic protons.  Yet,  the diffusive nature of the propagation is constrained  by the condition $R^2/6 D \gg R/c$. For the radius of the central molecular zone region of 200~pc,  this implies $D \ll  3  \times 10^{30} \ \rm cm^2/s$,  and, consequently,  $\dot{Q}_p  \ll  1.2 \times 10^{38} \ \rm  erg/s$. 
 
The energy spectrum of  the diffuse $\gamma$-ray emission (Fig. 3) has been extracted from an annulus centred at Sagittarius (Sgr) A* (see Fig. 1). The best-fit to the data is found for a spectrum following a power law extending with a  photon index $\approx$2.3 to energies up to tens of TeV, without any indication of a cutoff or  a break. This is the first time that such a $\gamma$-ray spectrum, arising from hadronic interactions, is detected in general. Since these $\gamma$-rays result from the decay of neutral pions produced by $pp$ interactions, the derivation of such hard power-law spectrum implies that the spectrum of the parent protons should extend to energies close to 1 PeV. The best fit of a $\gamma$-ray spectrum from neutral pion decay to the H.E.S.S. data is found for a proton spectrum following a pure power-law with index $\approx$2.4. We note that $pp$ interactions of 1 PeV protons could also be studied by the observation of emitted neutrinos or the X-rays from the synchrotron emission of secondary electrons and positrons (see Methods and Extended Data Figures 3 and 4).  However, the measured $\gamma$-ray flux puts the expected fluxes of neutrinos and X-rays below or at best close to the sensitivities of the current instruments. Assuming a cutoff in the parent proton spectrum, the corresponding secondary $\gamma$-ray spectrum deviates from the H.E.S.S. data at 68\%, 90\% and 95\% confidence levels for cutoffs at 2.9 PeV, 0.6 PeV and 0.4 PeV, respectively. This is the first  robust detection of a VHE cosmic hadronic accelerator which operates as a {\it PeVatron}. 

Remarkably, the  Galactic Centre PeVatron appears  to be located in the same region as the central 
$\gamma$-ray source  HESS J1745-290\cite{Aharonian:2004wa,2004ApJ...608L..97K,CANGAROO,MAGIC}. 
Unfortunately, the current data cannot provide an answer as to whether there is an intrinsic  link between these two objects.  The point-like source  HESS J1745-290 itself remains unidentified. Besides Sgr A*\cite{Aharonian:2005}, other potential counterparts are the pulsar 
wind nebula G 359.95-0.04\cite{wang2005,hinton06}, and a spike of annihilating 
dark matter\cite{belikov2012}. Moreover, it  has also been suggested that this source might have a diffuse origin, peaking towards the direction of the Galactic Centreg because of the higher concentration of both gas and relativistic particles\cite{Aharonian:2005}.  In fact, 
this interpretation would imply an extension of the spectrum of  the central source to energies beyond 10 TeV,  which however is at odds with the detection of a clear cutoff in the spectrum of HESS J1745-290 at about 10 TeV\cite{Aharonian:2009zk,VERITAS} (Fig. 3).   Yet, the attractive idea of explaining the entire  $\gamma$-ray emission  from  the Galactic Centre by run-away protons from the {\it same} centrally located accelerator  can be still 
compatible with the cutoff in the spectrum of the central source. For example, the cutoff could be due to the  absorption of $\gamma$-rays  caused by interactions with  the ambient infrared radiation field. It should be noted that although the question on the link between the central $\gamma$-ray source and the proton PeVatron  is an interesting issue in its own right, it, however, does not have a direct impact on the main 
conclusions of this work. 

The  integration  of  the cosmic-ray radial distribution (Fig.~2)  yields 
the  total energy of  E$\geq $10 TeV protons confined in the central molecular zone:   
$W_{\rm CR} \approx 1.0\times10^{49}$ erg. 
A single Supernova Remnant (SNR) would suffice to provide  this rather 
modest  energy in  cosmic rays. A  possible candidate could be Sgr A East. Although this object has already been excluded as a counterpart of HESS J1745-290\cite{Collaboration:2009tm}, the multi-TeV protons accelerated by this object and then injected into the central molecular zone could contribute
to the diffuse $\gamma$-ray component. Another potential site for acceleration of protons in the Galactic Centre  are the compact stellar 
clusters\cite{Crocker2011}. Formally, the mechanical power in these clusters in the form of  stellar winds, 
which  can provide adequate conditions  
for particle acceleration,  is   sufficient to explain the  
required total energy of cosmic rays in the central molecular zone.   However,  the 
acceleration of protons to PeV energies  
requires bulk motions  in excess of  10,000 km/s  which 
in the stellar clusters could only exist because of  
very young  supernova shocks\cite{Bykov}. Thus, the operation of  PeVatrons  in 
stellar clusters is reduced to the  presence of  supernovae shocks. 
On the other hand, since the acceleration of PeV particles by shocks, either 
in the individual SNRs or in the stellar clusters,  cannot last 
significantly longer than  100 years\cite{Bell:2013kq}, we would need more than 
10 supernova events  to meet the requirement of continuous injection of 
cosmic rays  in the central molecular zone over $\gg 10^3$ years.  For the central 10 pc region,  
 such a  high  supernova rate  seems  unlikely.    

We suggest  that  the supermassive black hole at the Galactic Centre (Sgr A*) is  the most plausible  supplier of ultra-relativistic protons and nuclei accelerated either in the accretion flow, \textit{i.e.} in the immediate vicinity of the black hole\cite{Aharonian:2005,2014NewA...27...13I} or somewhat further  away, e.g. at the site of termination of an outflow\cite{atoyan04}. If Sgr A* is indeed the particles' source, the required  acceleration rate  of about  $10^{37-38} $ erg/s  would  exceed by two or three orders of magnitude  the current bolometric luminosity  of 
Sgr  A*\cite{Genzel10},  and would constitute  at least 1 \% of the   current accretion power of  the supermassive black hole.    
Given the fact that the accretion rate in the central black hole  currently is relatively modest, and that at certain epochs the black hole of  $4 \times 10^6$ solar masses can operate at much higher accretion rate, we may speculate that this could facilitate also higher cosmic-ray production rates\cite{2014NewA...27...13I}.  In this regard it is interesting to note 
that  an average acceleration rate of $10^{39}$ erg/s of E$>$10 TeV protons over the last $10^{6-7}$ years would be sufficient to explain the flux of  cosmic rays around the so-called ``knee'' at 1~PeV. If so, this could be a solution to one of the most controversial  and actively debated problems of  the paradigm of the SNR origin of Galactic  cosmic rays\cite{Cristofari:2013pxd,Bell:2013kq,Parizot:2014ixa}.

\newpage

\begin{figure}[h!]
\includegraphics[width=1.0\textwidth]{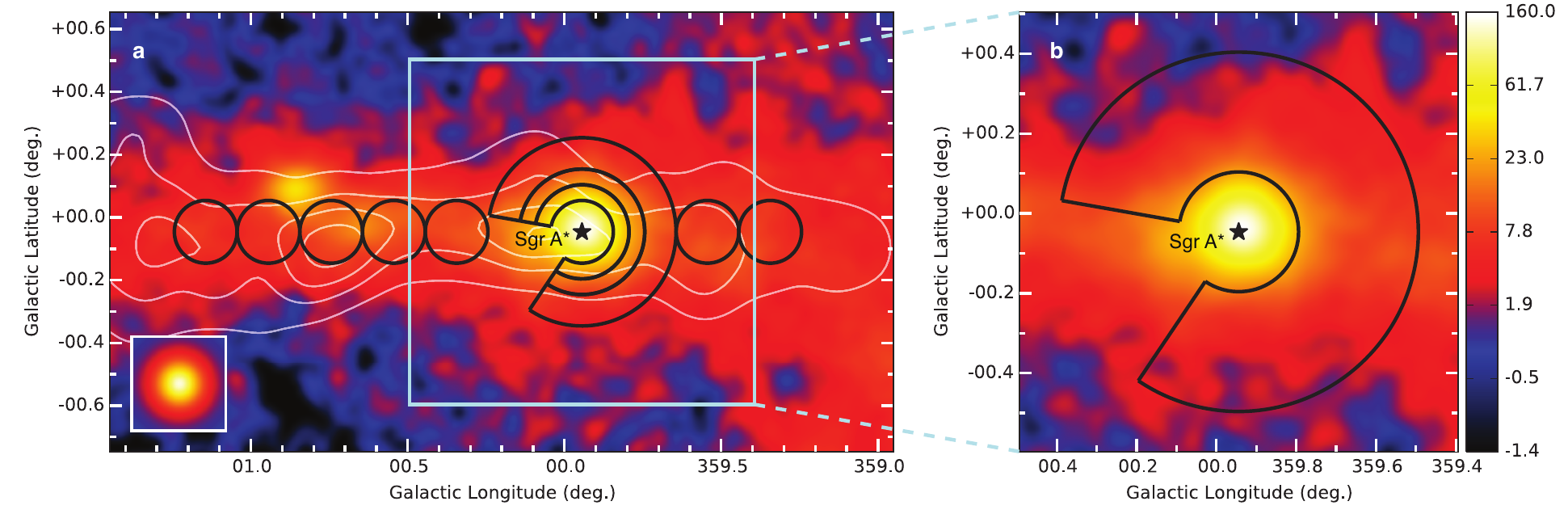} 
\caption{ {\bf VHE $\gamma$-ray image of the Galactic Centre region.} The colour scale indicates counts per 0.02$^{\circ}$$\times$0.02$^{\circ}$ pixel. \textit{Left panel:} The black lines outline the regions used to calculate the CR energy density throughout the central molecular zone. A section of 66$^{\circ}$ is excluded from the annuli (see Methods). White contour lines indicate the density distribution of molecular gas, as traced by its CS line emission\cite{tsuboi99}. The inset shows the simulation of a point-like source. \textit{Right panel:} Zoomed view of the inner $\sim 70$~pc and the contour of the region used to extract the spectrum of the diffuse emission.} 
\label{fig_map}
\end{figure}
\vfill

\begin{center}
\begin{figure}[th!]
\includegraphics[width=1.0\textwidth]{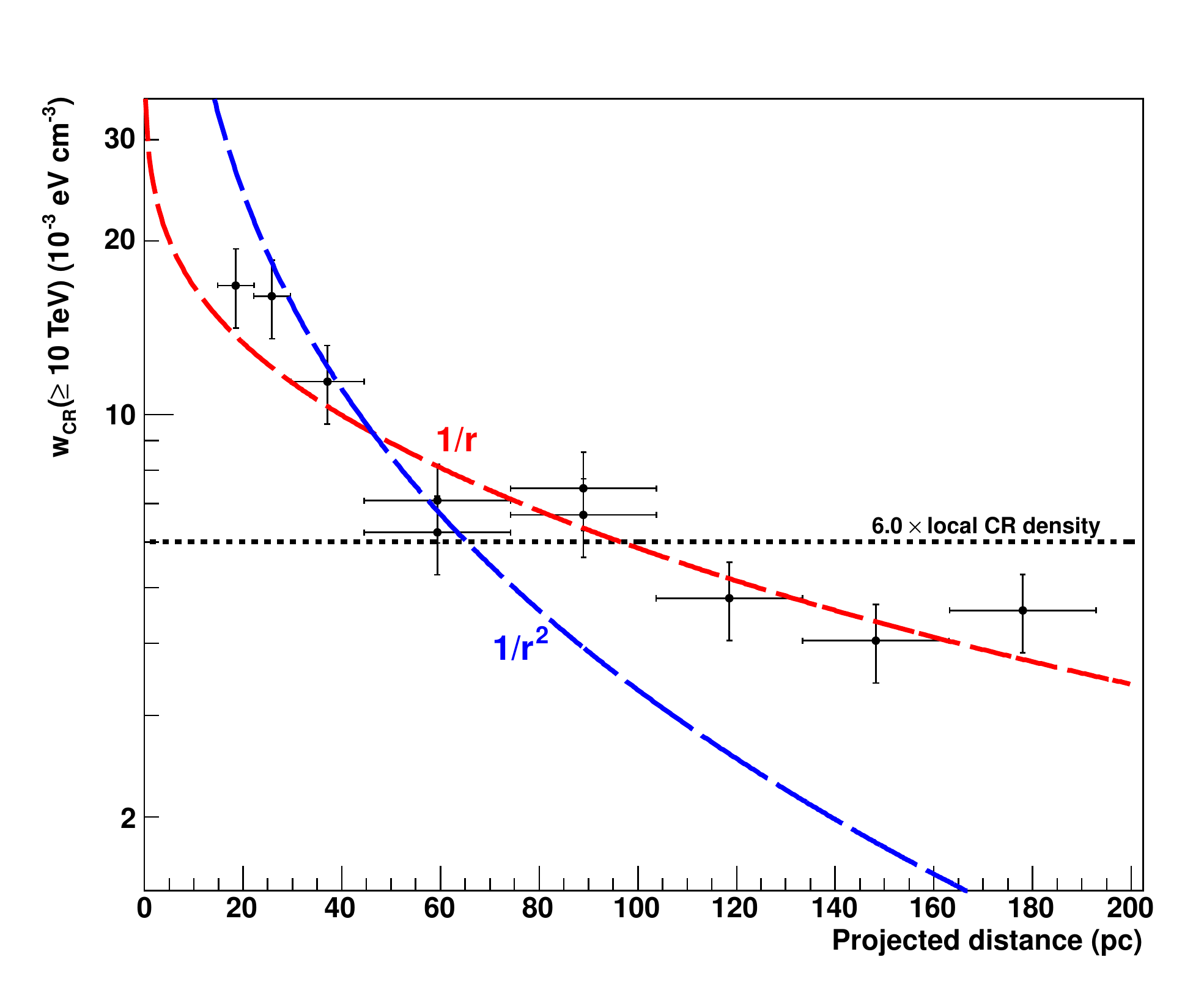}
\caption{{\bf Spatial distribution of the CR density versus projected distance from Sgr A*.} The vertical and horizontal error bars show the 1$\sigma$ statistical plus systematical errors and the bin size, respectively. A fit to the data of a $1/r$ (red line, $\chi^2/{\rm d.o.f.}$ = 11.8/9), $1/r^2$ (blue line, $\chi^2/{\rm d.o.f.}$ = 73.2/9) and an homogeneous (black line, $\chi^2/{\rm d.o.f.}$ = 61.2/9) CR density radial profiles integrated along the line of sight are shown. The best fit of a $1/r^\alpha$ profile to the data is found for $\alpha = 1.10 \pm 0.12$ (1$\sigma$).The $1/r$ radial profile is clearly preferred by the H.E.S.S. data.}
\label{fig_profile}
\end{figure}
\end{center}
\vfill
\begin{center}
\begin{figure}[]
\includegraphics[width=1.0\textwidth]{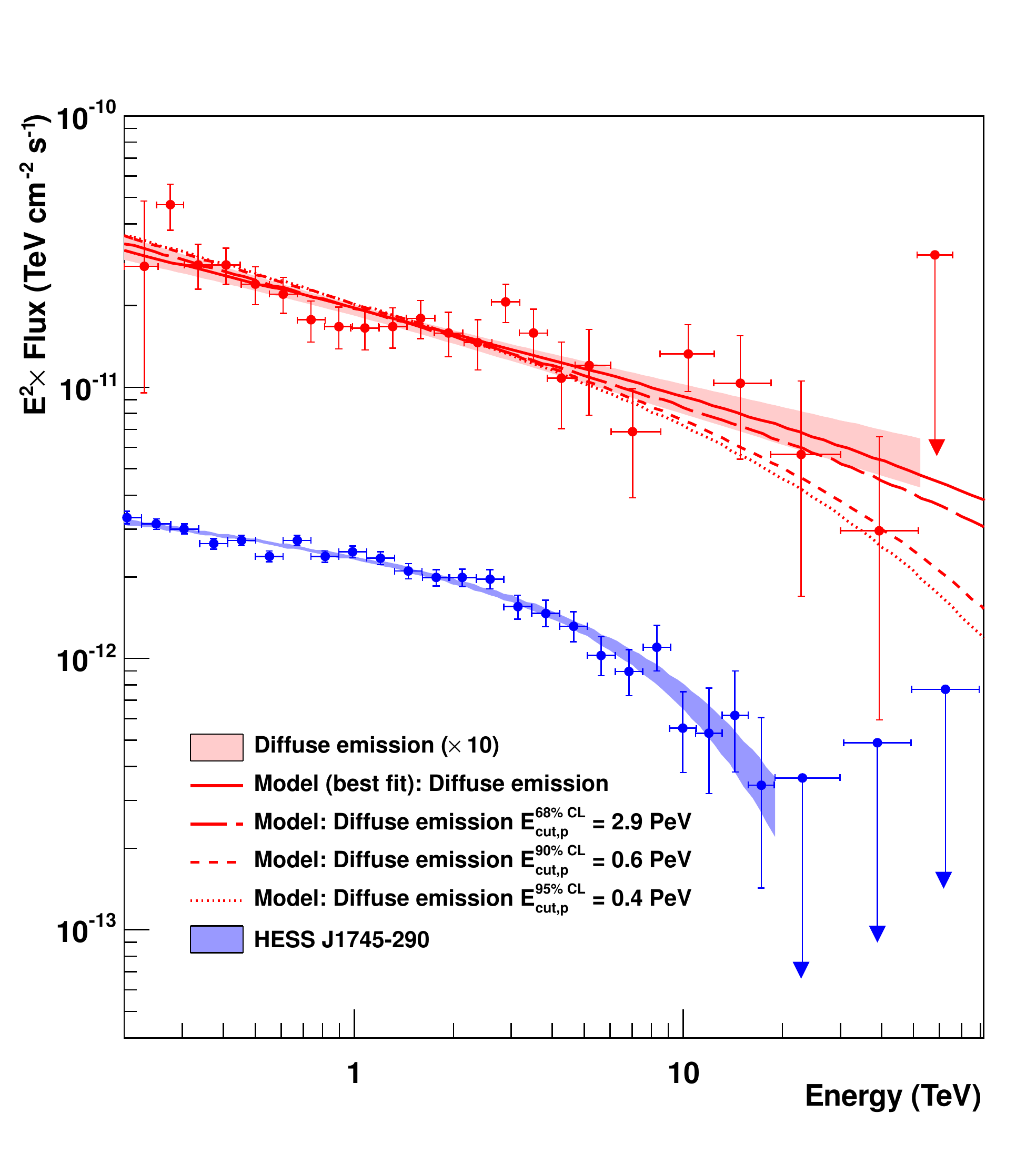}
\caption{{\bf VHE $\gamma$-ray spectra of the diffuse emission and HESS J1745-290.} The Y axis shows fluxes multiplied by a factor E$^2$, where E is the energy on the X axis, in units of TeVcm$^{-2}$s$^{-1}$. The vertical and horizontal error bars show the 1$\sigma$ statistical error and bin size, respectively. Arrows represent 2$\sigma$ flux upper limits. The 1$\sigma$ confidence bands of the best-fit spectra of the diffuse and HESS J1745-290 are shown in red and blue shaded areas, respectively. Spectral parameters are given in Methods. The red lines show the numerical computations assuming that $\gamma$-rays result from the decay of neutral pions produced by proton-proton interactions. The fluxes of the diffuse emission spectrum and models are multiplied by 10.}
\label{fig_spectrum}
\end{figure}
\end{center}
\vfill
\clearpage

\renewcommand{\refname}{References}

\newpage


\begin{addendum}

 \item The support of the Namibian authorities and of the University
of Namibia in facilitating the construction and
operation of H.E.S.S. is gratefully acknowledged, as is
the support by the German Ministry for Education and
Research (BMBF), the Max Planck Society, the German
Research Foundation (DFG), the French Ministry for
Research, the CNRS-IN2P3 and the Astroparticle Interdisciplinary
Programme of the CNRS, the U.K. Science
and Technology Facilities Council (STFC), the IPNP
of the Charles University, the Czech Science Foundation,
the Polish Ministry of Science and Higher Education,
the South African Department of Science and
Technology and National Research Foundation, and by
the University of Namibia. We appreciate the excellent
work of the technical support staff in Berlin, Durham,
Hamburg, Heidelberg, Palaiseau, Paris, Saclay, and in
Namibia in the construction and operation of the equipment.

\item[Author Contributions] F. Aharonian, S. Gabici, E. Moulin and A.Viana have analysed and interpreted the data, and prepared the manuscript. The whole H.E.S.S. collaboration has contributed to the publication with involvement at various stages ranging from the design, construction and operation of the instrument, to the development and maintenance of all software for data handling, data reduction and data analysis. All authors have reviewed, discussed, and commented on the present results and on the manuscript.

\item[Author Information] Reprints and permissions information is available at www.nature.com/reprints. The authors declare no competing financial interests. Correspondence and requests for materials should be addressed HESS Collaboration (contact.hess@hess-experiment.eu).

\end{addendum}

\newpage
\begin{center}
\bf{Methods}
\end{center}

\textbf{Data and spectral analyses} 

The phase-I of the H.E.S.S. array consists of four identical imaging atmospheric Cherenkov telescopes. The study presented in the Letter makes use of data collected from 2004 to 2013.  Data are taken in wobble mode where the pointing direction is chosen at an alternating offset of 0.7$^{\circ}$ to 1.1$^{\circ}$ from the target position\cite{Aharonian:2006pe}. Standard quality selection is applied to the data\cite{Aharonian:2006pe}, and after the selection procedure the dataset amounts to 226 hours of live time at the nominal position of Sgr A*. The analysis technique used to select the $\gamma$-ray events is based on a semi-analytical model of the air shower development\cite{deNaurois:2009ud}. The background level is calculated in each position using the Ring Background method\cite{Berge:2006ae}. The map shown in Fig.~1 is the $\gamma$-ray  excess count map per 0.02$\times$0.02 deg$^2$  smoothed by the H.E.S.S. point spread function and corrected for the telescope radial acceptance. 

The spectral reconstruction is based on a forward-folding method\cite{Piron:2001ir}. This method is based on a maximum-likelihood procedure, comparing the energy distributions of signal and background events to predefined spectral shapes. The energy spectra (Fig.~3) are both fitted by a power-law, $dN/dE = \Phi_1 (E/{\rm TeV})^{-\Gamma_1}$,  where $\Phi_1$ is the flux normalisation and $\Gamma_1$ is the spectral index, and by a power-law with an exponential cutoff, $dN/dE = \Phi_0 (E/{\rm TeV})^{-\Gamma_0}\times {\rm exp}(-E/E_{\rm cut})$, where $\Phi_0$ is the flux normalisation, $\Gamma_0$ is the spectral index and $E_{\rm cut}$ is the cut-off energy. A likelihood-ratio test between these two representations is performed to determine whether a significant deviation from a pure power-law is preferred by the data. The best-fit spectra together with their 1$\sigma$ confidence-level band are shown in Fig.~3 (shaded red and blue regions). The spectral points are uncorrelated flux points, obtained from the reconstructed best-fit spectrum. The error bars are the 1$\sigma$ Poisson deviation of the excess number of events in the energy bin. 

The diffuse emission spectrum is extracted from an annulus centered at Sgr A* (right panel of Fig. 1) with inner and outer radii of 0.15$^{\circ}$ and 0.45$^{\circ}$, respectively, and a solid angle of $1.4\times10^{-4}$ sr. The best-fit spectrum is given by a power-law with $\Phi_1 = (1.92 \pm 0.08_{\rm stat} \pm 0.28_{\rm syst}) \times$ $\rm 10^{-12}\,TeV^{-1}cm^{-2} s^{-1}$, and a photon index $\Gamma_1 =  2.32 \pm 0.05_{\rm stat} \pm 0.11_{\rm syst}$. Its $\gamma$-ray luminosity above 1 TeV is  $L_{\gamma} (\geq 1 {\rm TeV})= (5.69  \pm 0.22_{\rm stat} \pm 0.85_{\rm syst})\times 10^{34} {\rm erg \, s^{-1}}$. The fit of a power-law with an exponential energy cutoff is not preferred by the data. When compared with a pure power-law the likelihood-ratio test gives a p-value of 0.8 ($\sim$0.25 standard deviations from the power-law fit). In order to investigate the possibility of spatial variations of the spectral indices over the central molecular zone, the spectra within all the regions (left panel of Fig. 1) are reconstructed. All the indices are compatible, within a 1$\sigma$ standard deviation, to an index of $2.32 \pm 0.05_{\rm stat} \pm 0.10_{\rm syst}$  (see Extended Data Table 3). The spectrum of the central source is extracted from a circular region of radius 0.1$^{\circ}$ centered on Sgr A*. The best-fit spectrum is a power-law with an exponential cutoff with $\Phi_0 = (2.55\pm 0.04_{\rm stat} \pm 0.37_{\rm syst}) \times$ $\rm10^{-12}\,TeV^{-1}cm^{-2} s^{-1}$, a photon index of $\rm \Gamma_0 =  2.14 \pm 0.02_{\rm stat} \pm 0.10_{\rm syst}$, and an  energy cutoff at  $E_{\rm cut} = (10.7 \pm 2.0_{\rm stat} \pm 2.1_{\rm syst})$ TeV. When compared with a pure power-law the likelihood-ratio test gives a p-value of 3$\times$10$^{-5}$. A power-law with an exponential cutoff is clearly preferred by the data.

The diffuse $\gamma$-ray spectrum from the decay of neutral pions produced by $pp$ interactions reflects the parent proton spectrum. The normalisation of the proton spectrum depends on a combination of the injection power, target mass and propagation effects. Its shape, however, is completely defined by the observed $\gamma$-ray spectrum.  In the case of a $\gamma$-ray spectrum following a power-law with index $\Gamma_1$, the parent proton spectrum should follow a power-law with an index $\Gamma_{\rm p} \approx \Gamma_1 + 0.1$\cite{FA04}.  Using the parametrization of ref. 39, the best-fit proton spectrum to the H.E.S.S. data is obtained for a power-law with $\Gamma_{\rm p} \approx 2.4$. The corresponding 1$\sigma$ confidence band of this $\gamma$-ray spectrum is described by the red shaded area in Fig. 3. A fit to the data is also done with the $\gamma$-ray spectral shape derived from a proton spectrum following a power-law with an exponential cutoff. When compared with a power-law the likelihood-ratio test gives a p-value of 0.9 ($\sim$0.12 standard deviations from the power-law fit), and it is thus not preferred by the data. The lower limits on the proton spectrum energy cut-off can thus be derived from this fit. The 68\%, 90\% and 95\% confidence level deviation from the H.E.S.S. data is found for proton spectra following a power-law with $\Gamma_{\rm p} \approx 2.4$ and cutoffs at 2.9 PeV, 0.6 PeV and 0.4 PeV, respectively. Their corresponding $\gamma$-ray spectra are plotted in Fig. 3 (red long-dashed, dashed and dotted lines, respectively). The numerical computation of the energy spectrum of cosmic-ray protons with energy $>10$~TeV escaping from the central source at a rate of $\sim 8 \times 10^{37}$~erg/s over a time $\sim 6 \times 10^3$~yr, is shown in Fig. 3 (red solid line). Their injection spectrum is $Q(E) \propto E^{-2.2}$ extending up to 4 PeV, and their transport is described by a diffusion coefficient $D(E) = 6\times 10^{29} (E/10~{\rm TeV})^{\beta}$~cm$^2$/s, with $\beta = 0.3$. The resulting $\gamma$-ray spectrum is well compatible with the best-fit diffuse spectrum.

\textbf{Mass estimation and cosmic-ray energy density in the Central Molecular Zone}

The   derivation of the  cosmic-ray density  profile in the central molecular zone  rests  on  the  distribution of target material (for cosmic-ray interactions). The bulk of the gas in the Galactic Centre region is  in the form of the molecular hydrogen (H2), which is very difficult to detect directly. Therefore indirect methods of estimating the mass using tracer molecules must be applied. Tracer molecules are typically rare relative to H2 but much easier to detect and with an approximately known ratio to H2.  The mass estimates used in this paper are  based 
on the line emission of the CS molecule at the J=1-0 transition\cite{tsuboi99}. In order to evaluate the systematic uncertainties in these estimates,  other channels, such as  the line emission from transitions of  $^{12}$C$^{16}$O\cite{oka98} and HCN\cite{jones11} molecules are  also invoked.  The total mass in the inner 150 pc of the central molecular zone is estimated $(3^{+2}_{-1}) \times 10^{7}$M$_{\odot}$\cite{tsuboi99,Ferriere:2007yq}. The regions showed in Fig. 1 (left) almost completely cover the inner 150 pc of the central molecular zone and are used to extract the radial distribution of cosmic rays. They are symmetrically distributed around the centre of the central molecular zone, which is offset from Sgr A* by $\sim$50 pc in positive Galactic longitudes (l $\sim$ 0.33deg). These regions are described here below:

\begin{description}
\item[Ring 1 (R1) :]  Semi-annulus with [r$_{\rm in}$,r$_{\rm out}$]=[0.1$^{\circ}$,0.15$^{\circ}$], where r$_{\rm in}$ and r$_{\rm out}$ are the inner and outer radii, respectively. A section of 66$^{\circ}$ is excluded in order to avoid a newly detected source which will be reported elsewhere. This section is bounded by the opening angles of +10$^{\circ}$ and - 56$^{\circ}$  from the positive Galactic longitude axis (see Fig. 1). The average radial distance from Sgr A* is r$_{d} = 18.5$ pc.
\item[Ring 2 (R2) :] Semi-annulus  [r$_{\rm in}$,r$_{\rm out}$]=[0.15$^{\circ}$,0.2$^{\circ}$], r$_{d} = 25.9$ pc.
\item[Ring 3 (R3) :] Semi-annulus  [r$_{\rm in}$,r$_{\rm out}$]=[0.2$^{\circ}$,0.3$^{\circ}$], r$_{d} = 37.1$ pc.
\item[Circle 1/1b (C1/C1b) :]  circular region with 0.1$^{\circ}$ of radius centred at l = 0.344$^{\circ}$/359.544$^{\circ}$, b = $-0.04588^{\circ}$, in galactic coordinates, and r$_{d}$ = 59.3 pc
\item[Circle 2/2b (C2/C2b):]  circular region with 0.1$^{\circ}$ of radius centred at l = 0.544$^{\circ}$/359.344$^{\circ}$, b = $-0.04588^{\circ}$, r$_{d}$ = 89.0 pc
\item[Circle 3 (C3):]  circular region with 0.1$^{\circ}$ of radius centred at l = 0.744$^{\circ}$, b = $-0.04588^{\circ}$, r$_{d}$ = 118.6 pc
\item[Circle 4 (C4):]  circular region with 0.1$^{\circ}$ of radius centred at l = 0.944$^{\circ}$, b = $-0.04588^{\circ}$, r$_{d}$ = 148.3 pc
\item[Circle 5 (C5):]  circular region with 0.1$^{\circ}$ of radius centred at l = 1.144$^{\circ}$, b = $-0.04588^{\circ}$, r$_{d}$ = 178.0 pc
\end{description}

If the $\gamma$-ray emission is completely due to the decay of neutral pions produced in  proton-proton interactions, then the $\gamma$-ray luminosity $L_{\gamma}$ above the  energy $E_{\gamma}$ is related to the total energy of cosmic-ray protons  $W_{p}$ as 
\begin{equation}
L_{\gamma} (\geq E_{\gamma} ) \sim \eta_N \frac{ W_{p} (\geq 10 E_{\gamma}) }{ t_{pp \rightarrow \pi^0} } \, ,
\end{equation}
where $t_{pp \rightarrow \pi^0} = 1.6 \times 10^{8}$ yrs $({\rm 1 cm}^{-3}/n_{\rm H} )$ is the proton energy loss timescale due to neutral pion production in an environment of hydrogen gas of  density $n_{\rm H}$\cite{FA04}, and $\eta_N \approx 1.5$ 
accounts for the presence of nuclei heavier than hydrogen in both cosmic rays and interstellar matter.   
The energy density of cosmic rays $w_{\rm CR}$, averaged along the line of sight is then:
\begin{equation}
w_{\rm CR} (\geq 10 E_{\gamma}) = \frac{W_p(\ge 10 E_{\gamma})}{V} \sim 1.8 \times 10^{-2} \left( \frac{\eta_N}{1.5} \right)^{-1} \left( \frac{L_{\gamma} (\geq E_{\gamma} )}{10^{34} {\rm erg/s}} \right) \left( \frac{M}{10^6 M_{\odot}} \right)^{-1} \rm eV/cm^3 \, ,
\end{equation}
where $M$ is the mass of the  relevant region. The $\gamma$-ray luminosity above 1 TeV and the mass estimates 
(based on three tracers) for all regions  are presented in Extended Data Table~1.  The cosmic-ray energy densities in different regions, given in units of  $10^{-3}$ eV/cm$^{3}$, which is the value of the local cosmic-ray energy density $w_{\rm 0}{(\geq 10 {\rm TeV})}$ (as measured in the Solar neighbourhood), are presented in Extended Data Table~2.

The uncertainty in  the cosmic-ray energy density comes basically from the uncertainty in the mass estimates. The independent estimates from different tracers result in the cosmic-ray enhancement factors in the inner regions of the central molecular zone ($r \leq$ 25 pc from the Galactic Centre) 16$_{-5}^{+10}$ (CS),  22$_{-7}^{+14}$ (CO) and 24$_{-8}^{+16}$ (HCN). The cosmic-ray radial distribution is also computed for all the different channels.  The results  (shown in Fig.2 for CS line tracer) are in good agreement with the $1/r$ profile with $\chi^2/{\rm d.o.f.}$ = 11.8/9 , $\chi^2/{\rm d.o.f.}$ = 9.4/9 and $\chi^2/{\rm d.o.f.}$ = 11.0/8 for mass estimates based on the CS, CO and HCN tracers, respectively.  At the same time, the data are in obvious conflict with the $w(r)  \propto 1/r^2$, $\chi^2/{\rm d.o.f.}$ = 73.2/9 (CS), $\chi^2/{\rm d.o.f.}$ = 78.0/9 (CO) and $\chi^2/{\rm d.o.f.}$ = 57.9/8 (HCN), and $w(r) \propto constant$ profiles, $\chi^2/{\rm d.o.f.}$ = 61.2/9 (CS), $\chi^2/{\rm d.o.f.}$ = 45.6/9 (CO) and $\chi^2/{\rm d.o.f.}$ = 77.1/8 (HCN). Finally, when fitting a $1/r^{\alpha}$ profile to the data, the best fit is found for $\alpha$ equal to $1.10 \pm 0.12$ (CS), $0.97 \pm 0.13$ (CO) and $1.24 \pm 0.12$ (HCN), with $\chi^2/{\rm d.o.f.}$ = 11.1/8, $\chi^2/{\rm d.o.f.}$ = 9.34/8 and $\chi^2/{\rm d.o.f.}$ = 6.5/7, respectively, which confirms the preference for an $1/r$ density profile to describe the data.  \\

\textbf{Spectral analysis within the central molecular zone}

The findings of a PeVatron accelerating protons in a quasi-continuous regime, over a sufficiently long period of time to fill the whole central molecular zone, implies that the gamma-ray energy spectral shape should be spatially independent over the central molecular zone. 
The available statistics in each of these regions prevent us to test the existence of a cutoff beyond 10 TeV. The indices of all the regions are presented in Extended Data Table~3. All the indices are compatible within a 1$\sigma$ standard deviation to an index of $2.32 \pm 0.05_{\rm stat} \pm 0.10_{\rm syst}$, as measured in the annulus in the right panel of Fig. 1, and used to derive the properties of the Galactic Centre PeVatron.  The compatibility of the measured spectral indices over the 200 pc of the central molecular zone provides an additional piece of evidence for the scenario proposed in this Letter.\\

\textbf{Multi-TeV $\gamma$-rays of  leptonic origin?}
 
For the  diffuse  $\gamma$-ray emission of  Galactic Centre   we deal  with quite  specific conditions which 
strongly constrain the possible scenarios of $\gamma$-ray production.  Two major radiation 
mechanisms are related to interactions of ultrarelativistic 
protons and electrons, with the dense gas in the central molecular zone and with the ambient  infrared radiation fields, respectively. 
For explanation of multi-TeV $\gamma$-rays, the maximum energy of protons and electrons should be as large as $\sim$1~PeV and $\sim$100~TeV, respectively.  Additionally, these particles should  effectively propagate and fill  the entire central molecular zone.   While in the case of the  hadronic  scenario 
one needs  to postulate an existence of a PeVatron in the Galactic Centre,  any `leptonic' model of $\gamma$-ray production  should address  the following   questions: 
(i) whether the accelerator could be sufficiently effective to boost the energy of electrons up to
$\geq \ 100$~TeV  under the severe radiative losses in the Galactic Centre; (2) whether these electrons can escape the sites of their production and propagate over distances of tens of parsecs; (3) whether they can explain the  observed hard spectrum of multi-TeV $\gamma$-rays. 

Acceleration of electrons to multi-100 TeV energies is  more difficult than acceleration 
of protons because of severe synchrotron and inverse Compton (IC) losses. Formally, acceleration of electrons  to energies beyond 100~TeV is possible in the so-called extreme accelerators, where the acceleration proceeds at the maximum possible rate allowed by classical electrodynamics, 
$t_{\rm acc}  \sim R_{\rm L}/c  \approx 0.4 (E/100 \ \rm TeV) (B/1 \ \mu G)^{-1}$~yr, where $R_{\rm L}$ is the Larmor radius.   Even 
so, the escape of such energetic electrons from the accelerator and  their propagation far enough (tens of parsecs) to fill the central molecular zone, can be realised only for rather unrealistically weak magnetic fields and  fast diffusion. Indeed, the propagation time over a distance $R$ and for a particle diffusion coefficient $D$ is equal to $t_{\rm diff} = R^2/6 D \sim 2 \times 10^3 (R/200~{\rm pc})^2 (D/10^{30}~{\rm cm^2/s})$~yr and, for typical interstellar conditions, is much 
longer than the synchrotron loss time of electrons with energy $E_e$, 
$t_{\rm synch} \approx 10 (B/100 \mu \rm G)^{-2} (E_e/100 \ \rm)^{-1} \ \rm yr$.

The  efficiency  of a given $\gamma$-ray emitting process  is determined by the cooling time of particles  through that specific channel  compared to  the characteristic times of other 
(radiative and non-radiative) processes.  The cooling times of relativistic electrons in the central molecular zone  
are shown in Extended Data Fig.~1.   While bremsstrahlung and IC scattering  result in $\gamma$-ray emission,  the ionisation and synchrotron losses  reduce the efficiency of $\gamma$-ray production. Bremsstrahlung is an effective mechanism of $\gamma$-radiation at GeV energies.  Above 100 GeV  the IC cooling becomes more effective ($t_{\rm IC} < t_{\rm br}$; where $t_{\rm IC}$ and $t_{\rm br}$ are the electrons' cooling times through IC and bremsstrahlung, respectively), and strongly dominates over bremsstrahlung at energies above 10 TeV.  This can be seen in Extended Data Fig.~2 where  the results of calculations of the spectral energy distribution of  broad-band emission of electrons  are shown.  The calculations are performed for an acceleration spectrum following  a power-law with an exponential cutoff at 100 TeV.  Assuming that electrons are injected in a continuous regime, the steady-state spectrum of electrons is obtained by solving the kinetic equation which takes into account the energy losses of electrons due to ionisation, bremsstrahlung, synchrotron radiation and IC scattering. At low energies more important are the losses  due to the diffusive escape of electrons from the central molecular zone.  Although it has been shown that the magnetic field in the Galactic Centre should have a lower-limit of  $B=50 \, \mu$G on 400 pc scales\cite{2010Natur.463...65C}, here we assume a very low $B=15 \, \mu$G. Even with such low magnetic field, it is seen that above 10 TeV the calculations  do not match the observed fluxes. Formally, one can assume, higher, e.g. by an order of magnitude,  gas density (e.g. if $\gamma$-rays are produced mainly in  dense cores of molecular clouds), thus the bremsstrahlung would dominate over the IC contribution, and the flux of $\gamma$-rays could be increased.  However, for  any reasonable magnetic field, the synchrotron losses above 10 TeV will dominate over bremsstrahlung. This will make  the  steady-state electron spectrum steeper with power-law index $\alpha=\alpha_0+1$ ($\alpha_0$ is the power-law index of the electron  injection spectrum).  Since the $\gamma$-ray spectrum produced due to bremsstrahlung mimics the energy spectrum of parent electrons ($\Gamma=\alpha$), at energies of $\gamma$-rays above a few TeV we should expect quite steep  spectrum of $\gamma$-rays with a power-law index $\Gamma > 3.4$. This is in apparent conflict with observations. \\

\textbf{Multiwavelength and multi-messenger signatures of PeVatrons}

Galactic PeVatrons have unique signatures  which  allow  their  unambiguous identification among other particle accelerators.  Such signatures are  related to the neutral secondary products resulting from hadronic interactions of accelerated PeV protons and nuclei ($E \geq 0.1$ PeV/nucleon) with the ambient gas. The secondaries  produced at low energies, in particular MeV/GeV $\gamma$-rays and the radio synchrotron emission of primary and secondary electrons and positrons, do carry information about the accelerator, however, strictly speaking, they are not directly linked to the PeV particles. The extrapolations from low  to high energies based on theoretical assumptions are model-dependent, and therefore biased. Obviously, they cannot substitute the direct measurements at highest energies.

All three products of  interactions of ultrarelativistic protons -  $\gamma$-rays, neutrinos and electrons - generated  through the production and decay of $\pi^0, \pi^+$ and $\pi^-$ mesons, receive  approximately  10\% of  the energy of  primary  protons, thus  multi-TeV  secondary neutrals carry  univocal information about the  primary PeV protons.  This concerns, first of all,   $\geq 10$~TeV $\gamma$-rays,  because  at such high energies  the efficiency of leptonic channels of production of high energy $\gamma$-rays in general, and 
in the central molecular zone, in particular, is dramatically reduced (see above). 
The flux sensitivity as well as the  angular and energy resolutions  achieved by  the H.E.S.S. array  allow  adequate studies of  the acceleration  sites and the  propagation  of  accelerated protons up to  1~PeV  based on the  morphological and spectral  properties of multi-TeV $\gamma$-rays from the Galactic Centre. 

An independent and straightforward proof of the hadronic origin of diffuse $\gamma$-rays from the  central molecular zone  would be the detection of multi-TeV neutrinos spatially correlated with $\gamma$-rays. In Extended Data Fig.~3 we  show  the fluxes of high energy neutrinos  which should accompany  the $\gamma$-ray flux  presented in Fig.~3. The calculations are based on  the parent proton spectrum derived from  $\gamma$-ray data, therefore the only free parameter  in these calculations is the high energy cutoff $E_0$ in the spectrum of parent protons. The  condition for the detection  of  high energy neutrinos by km$^3$-scale detectors (such as IceCube or KM3Net) can be expressed by a minimum flux of  $\gamma$-rays, assuming that both neutrinos and $\gamma$-rays are products of $pp$ interactions.  
The estimate of detectability of neutrinos is most robust (i.e.  less sensitive to the spectral shape) when 
normalised at a particle energy of $\approx 20$ TeV. Namely, neutrinos can be detected by a km$^3$ volume  detector if the  differential flux of accompanying $\gamma$-rays  at 20~TeV  exceeds $10^{-12}$ TeV cm$^{-2}$ s$^{-1}$  \cite{vissani}.   Since the $\gamma$-ray fluxes (Fig.~3) are quite close to this value, we may conclude that  multi-TeV neutrinos from the Galactic Centre can be  marginally 
detected by a km$^3$-scale detector after several years of exposure.  

The third  complementary  channel of information about the PeV protons is carried by   the   secondary electrons, through their synchrotron radiation. In $pp$ interactions,  electrons  are produced in a fair balance with  neutrinos and photons 
(their distribution almost coincides with the spectrum of the electronic neutrinos)\cite{kelner},  {\it i.e.} they carry a significant fraction of the energy of the incident proton. In  environments  with magnetic field $B \geq 100 \  \mu \rm G$, the  lifetime of secondary electrons  producing X-rays of energy $\epsilon_x$ is quite short, $\sim 15 (B/100 \  \mu \rm G)^{-3/2} 
(\epsilon_x/10 \rm \ keV)^{-1/2}$~yr, compared to 
other characteristic times, e.g.  to the propagation time of protons over the central molecular zone zone. 
Therefore,  
hard X-rays can be considered as a ``prompt'' radiation component emitted in hadronic interactions simultaneously with $\gamma$-rays and neutrinos.  The energy release in X-rays  calculated self-consistently  for the proton spectrum derived from the $\gamma$-ray data,     
exceeds 10 \% of the energy released  in  multi-TeV neutrinos and $\gamma$-rays 
(see in Extended Data Fig.~4).  In general, this is quite substantial, given the  superior flux sensitivity of X-ray instruments,  especially for point like sources. However,  since the radiation component shown in Extended Data Fig.~4  is integrated over a very large (for X-ray instruments) region with an angular size of $\sim  1^\circ$, it is  overshadowed  by the diffuse X-ray emission detected by XMM-Newton\cite{2013MNRAS.428.3462H}. This makes  the detection of this component practically impossible. 

Finally,  one should mention  relativistic  neutrons as another potential 
messengers of hadronic processes produced in the  Galactic Centre  predominantly in the reaction 
$p  p \rightarrow  p n  \pi^+$.  However, because of the short 
lifetime of about  $\tau \simeq 10^3$~sec, only neutrons of energy $E_{\rm n} \geq m_{\rm n}  d/(c \tau) \sim 10^{18}$eV (where $m_{\rm n}$ is the neutron mass) could reach us 
before they decay during the  ``free flight"  from the Galactic Centre ($d=8.5$~kpc is the distance to the Galactic Centre). 
Thus if  the proton spectrum  in the Galactic Centre extends to extremely high energies, 
one can probe, in principle, this additional channel of information by  detectors of cosmic rays, 
in particular by the AUGER observatory.  \\

\textbf{Alternative explanations to the supermassive black hole as the source of cosmic rays responsible for the diffuse $\gamma$-ray emission in the central molecular zone}

Several cosmic-ray sources are present in the Galactic centre region. Besides the Galactic Centre supermassive black hole, discussed 
in the Letter, alternative sources of the cosmic rays responsible for the $\gamma$-ray emission from the central molecular zone include supernova remnants (SNRs)\cite{okkie}, stellar clusters\cite{Crocker2011}, and radio filaments\cite{farad}.
Any of these scenarios  should satisfy the following conditions, derived from the H.E.S.S. observations of the central molecular zone: {\it i)} the accelerator has to be located in the inner $\sim$10 pc of the Galaxy, {\it ii)} the accelerator(s) has(have) to be continuous over a timescale of at least thousands of years, and {\it iii)} the acceleration has to proceed up to PeV energies.

{\it Supernova remnants --} The acceleration mechanism operating at SNRs is widely believed to be diffusive shock acceleration, characterized by an acceleration timescale $t_{acc} \sim D(E)/u_s^2$, where $D(E) \propto E/B$ is the Bohm diffusion coefficient of a cosmic ray of energy $E$ in  a magnetic field $B$ and $u_s$ is the shock speed. It is clear, then, that the fastest acceleration rate is obtained for the largest possible values of the shock speed and of the magnetic field strength.

In the early free expansion phase of the SNR evolution, shock speeds as high as $\approx 10^4$~km/s can be achieved. The magnetic field strength is also expected to be very large during this early phase (up to $\sim 0.1-1$~mG\cite{vink}) due to efficient field amplification connected to the acceleration of cosmic rays at the shock \cite{bell04}. Under these circumstances, SNRs can accelerate protons up to an energy of $E_{max} \approx 10^{14} ( B/100~\mu{\rm G}) ( u_s/10000 ~ {\rm km\, s^{-1}} )^2 ( \Delta t_{\rm PeV}/\rm yr)  ~ \rm eV$, where $\Delta t_{\rm PeV}$ is the duration of the phase where shock speed and magnetic field are high enough to allow the acceleration of particles up to PeV energies. It has been shown in ref. 24 that the duration of this phase is of the order of tens of years (definitely less than a century). Thus, even though SNRs can potentially provide PeV particles, they cannot act as (quasi) continuous injectors of such energetic particles for a time of the order of thousands of years.

{\it Stellar clusters --} Other possible cosmic-ray sources in the Galactic Centre region are stellar clusters. Three of them are known in the inner $\sim 0.1^{\circ}-0.2^{\circ}$ region: the central, the Arches, and the Quintuplet cluster. The most likely sites for acceleration of particles in stellar clusters are the stellar winds of the massive OB stars that form the cluster, and the shocks of the supernovae which mark the end of the life of these stars. However, in order to accelerate particles up to PeV energies, very large shock velocities, of the order of at least 10000 km/s are needed \cite{Bell:2013kq}. Velocities of this order can hardly be found in stellar wind termination shocks, and thus  SNR shocks following the explosion of cluster member stars remain the best candidates as particle accelerators.

Both the Arches and the Quintuplet clusters are located outside of the inner $\sim 10$~pc region\cite{Bykov}, and this disfavours their role as accelerators of the cosmic rays responsible for the diffuse $\gamma$-ray emission.  On the contrary, the central cluster is located well within the central 10 pc region, and thus should be considered as a potential candidate for the acceleration of cosmic rays in the central molecular zone. The $\gamma$-ray observations 
suggest that the cosmic-ray source  in the Galactic centre  should act (quasi) continuously over  time $t_{\rm inj}$ of a few thousands of years. Given that a SNR can accelerate PeV particles over a time interval $t_{\rm PeV}$ of less than a century, we would need at least $\sim  10 (t_{\rm PeV}/100 {\, \rm yr}) (t_{\rm inj}/1000 {\, \rm yr})^{-1}$ supernova explosions happening over the last $t_{\rm inj}$ within the central cluster. 
Given the very small size of the region  ($\sim$ 0.4 pc), such a large SN explosion rate is unrealistic. 

{\it Radio filaments --} It has been proposed in ref. 45 that the diffuse $\gamma$-ray emission from the central molecular zone  was  the  result of non-thermal Bremsstrahlung from relativistic electrons\cite{farad,farad2}. In this scenario, the putative sources of $\gamma$-ray emitting electrons are the elongated radio filaments detected throughout the central molecular zone region\cite{farad}. This is in tension with one of the main finding of this paper, i.e. the location of the source of cosmic rays in the inner $\sim 10$~pc of the Galaxy. Though the acceleration mechanism is not discussed, filaments are assumed to somehow accelerate electrons and then release them in the interstellar medium. In order to fill the whole central molecular zone region before being cooled by synchrotron and inverse Compton losses, electrons are assumed to propagate ballistically (i.e. at the speed of light  without  a significant deflection in the magnetic field). This  unconventional assumption is made at the expenses of a very large energy requirement: observations can be explained if the energy injection rate of cosmic-ray electrons in the central molecular zone is of the order of $\sim 10^{41}$~erg/s\cite{farad}. This is a very large injection rate, being comparable to the total luminosity of cosmic-ray protons in the whole Galaxy, and makes this scenario problematic.\\\\

\renewcommand{\figurename}{Extended Data Figure}
\setcounter{figure}{0}   

\begin{figure}[]
\includegraphics[width=1.0\textwidth]{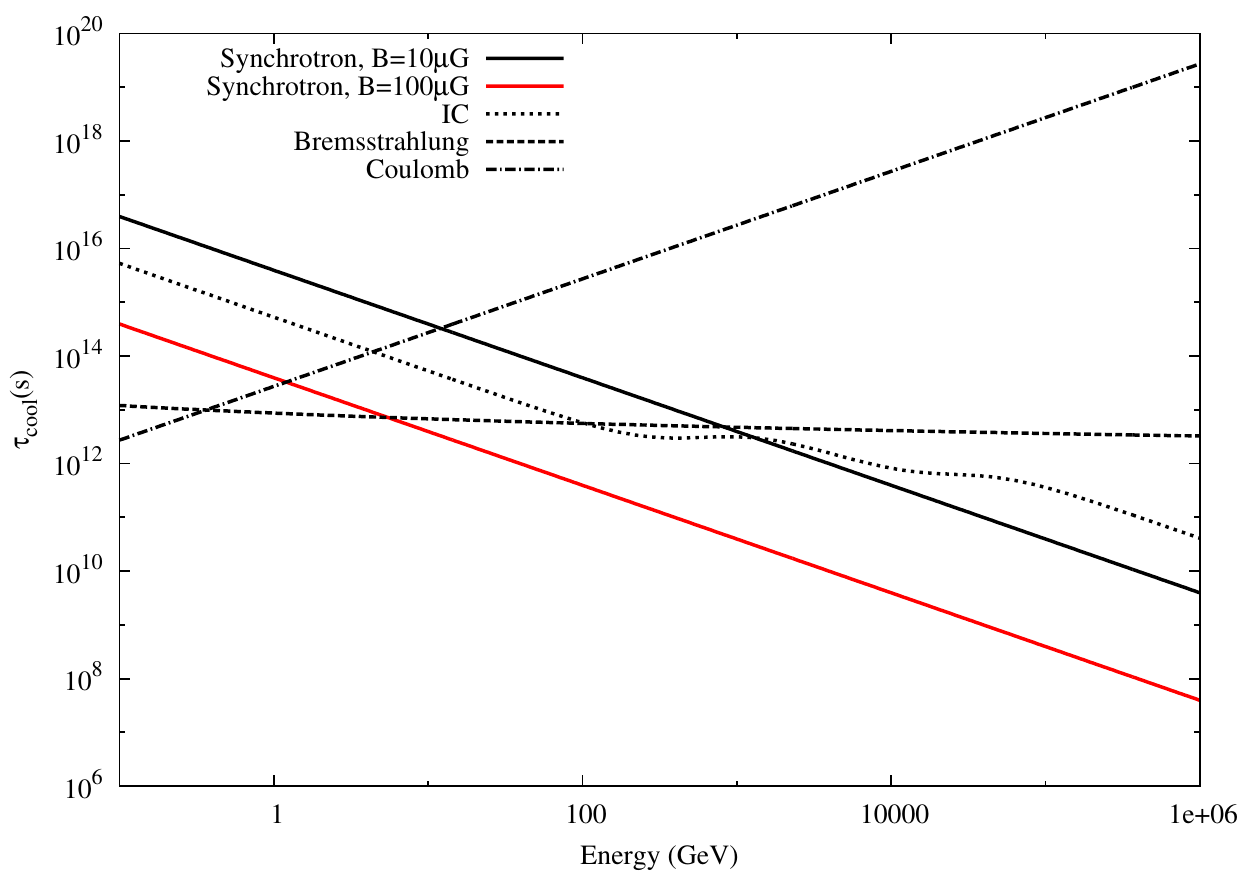} 
\caption{{\bf Cooling times of electrons  in the Galactic Centre as a function of energy.}  The cooling times due to ionisation (or Coulomb) losses  and bremsstrahlung are inversely proportional to the gas density $n$;  here $n=100 \rm \,cm^{-3}$ is assumed.  The cooling time of the synchrotron radiation is proportional to $1/B^2$, where B is the magnetic field. The total energy densities of the Cosmic Microwave Background, local near (NIR) and far (FIR) infrared radiation fields used to calculated the cooling time due to the IC scattering are extracted from the GALPROP code\cite{1998ApJ...509..212S}. The integrated densities are 17.0 and 1.3 $\rm eV/cm^{-3}$ for NIR and FIR, respectively.}
\label{neutrino}
\end{figure}
\vfill
\clearpage
\begin{figure}[]
\includegraphics[width=1.0\textwidth]{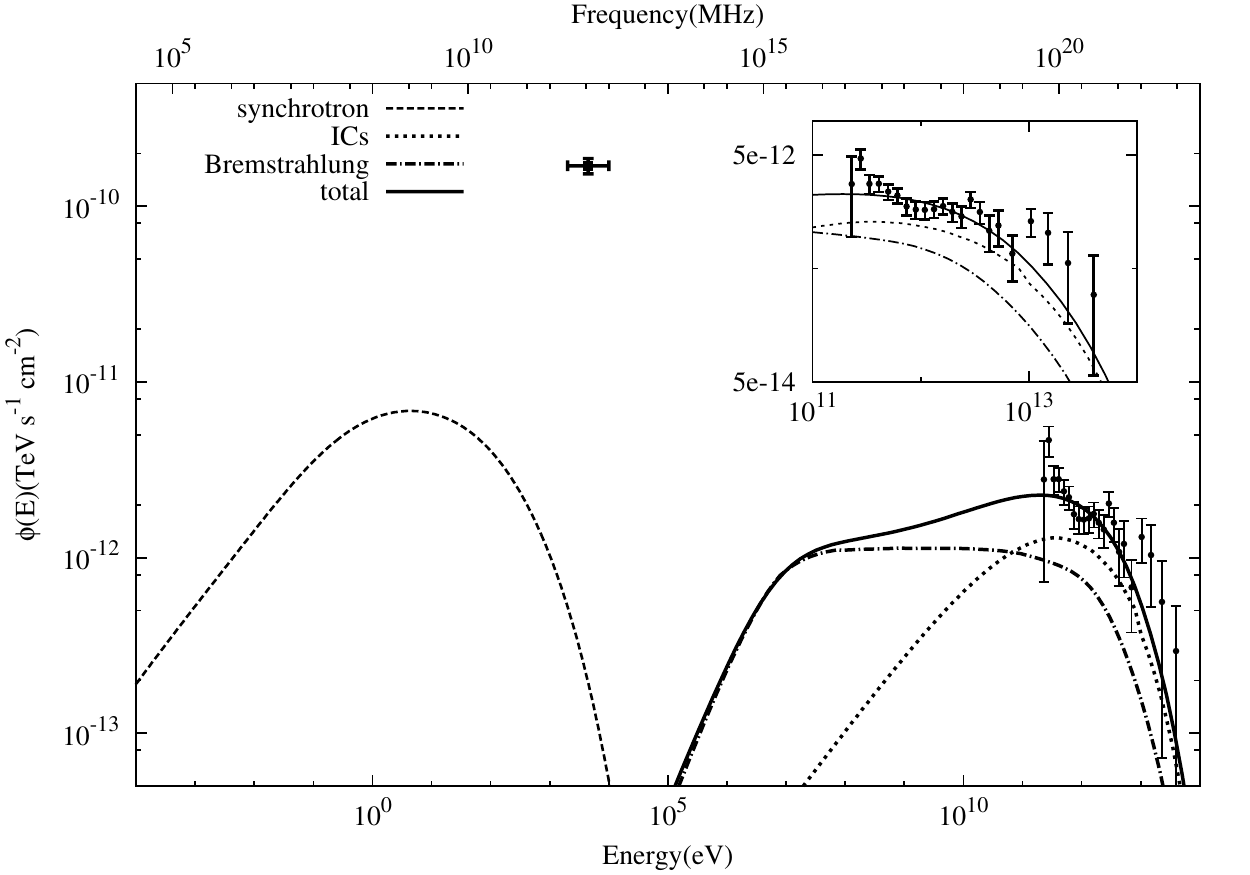} 
\caption{{\bf Broad-band spectral energy distribution of radiation by relativistic 
electrons.} The flux from synchrotron radiation, bremsstrahlung and the inverse 
Compton scattering are compared to the fluxes of diffuse $\gamma$-ray emission measured by H.E.S.S. (black points with vertical error bars). The flux of  diffuse X-ray emission  measured by XMM-Newton\cite{2013MNRAS.428.3462H}  (black point with horizontal error bar) and integrated  over the  central molecular zone region is also shown. The inset (top right) shows a zoomed view of the SED at the VHE range (100 GeV - 100 TeV). The vertical and horizontal error bars show the 1$\sigma$ statistical errors and the bin size, respectively.$\mu$G.  The flux of  diffuse X-ray emission  measured by XMM-Newton and integrated  over the  central molecular zone region   is also shown. }
\label{Xrays}
\end{figure}
\vfill
\clearpage

\begin{figure}[]
\includegraphics[width=1.0\textwidth]{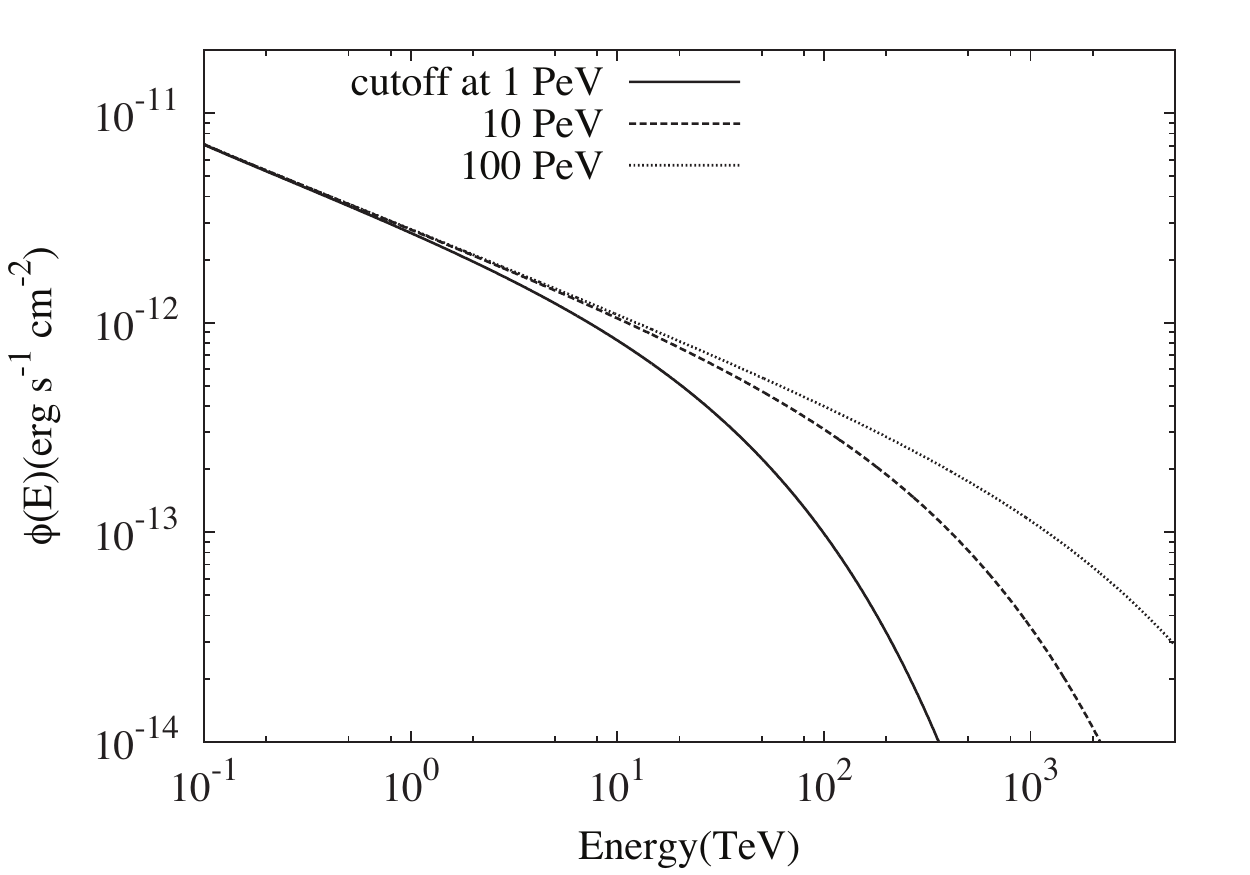} 
\caption{{\bf The spectral energy distribution of  high energy neutrinos - the counterparts 
of diffuse $\gamma$-rays from the Galactic Centre.}  The energy spectrum of parent protons is  derived from the $\gamma$-ray data.  The three curves correspond to different values of the  exponential cutoff in the proton spectrum: 1 PeV, 10 PeV, 100 PeV.}
\label{cooling}
\end{figure}
\vfill
\clearpage

 \begin{figure}[]
\includegraphics[width=1.0\textwidth]{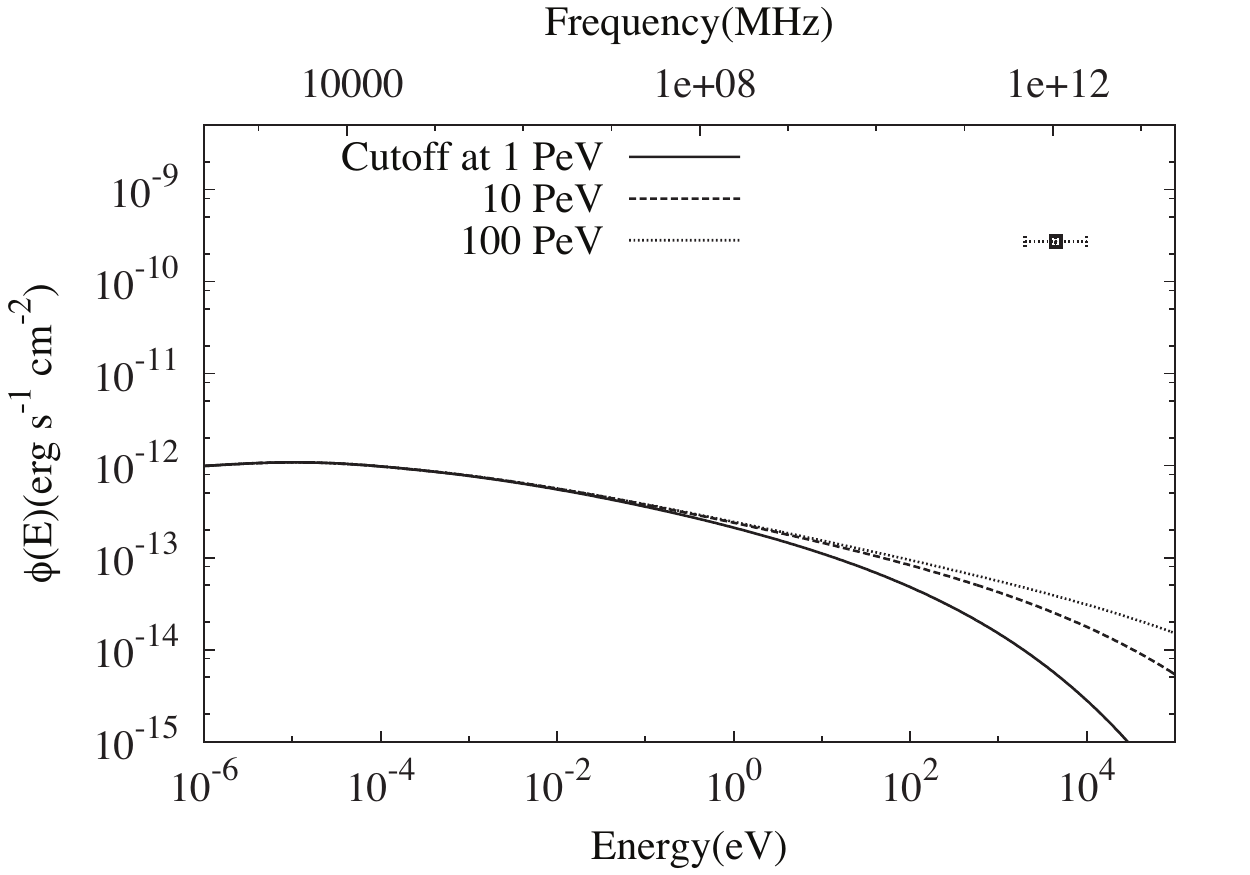} 
\caption{{\bf The spectral energy distribution of synchrotron radiation of 
secondary electrons produced in $pp$ interactions.}  The spectra of  protons are the same as in 
Extended Data Fig.~3.  The magnetic field is assumed 100~$\mu$G.  The flux of  diffuse X-ray
emission  measured by XMM-Newton and integrated  over the  central molecular zone region   is also shown. The horizontal error bar correspond to the bin size.}
\label{SED}
\end{figure}
\vfill
\clearpage

\renewcommand{\figurename}{Extended Data Table}
\setcounter{figure}{0}   

 \begin{figure}[]
\includegraphics[width=1.0\textwidth]{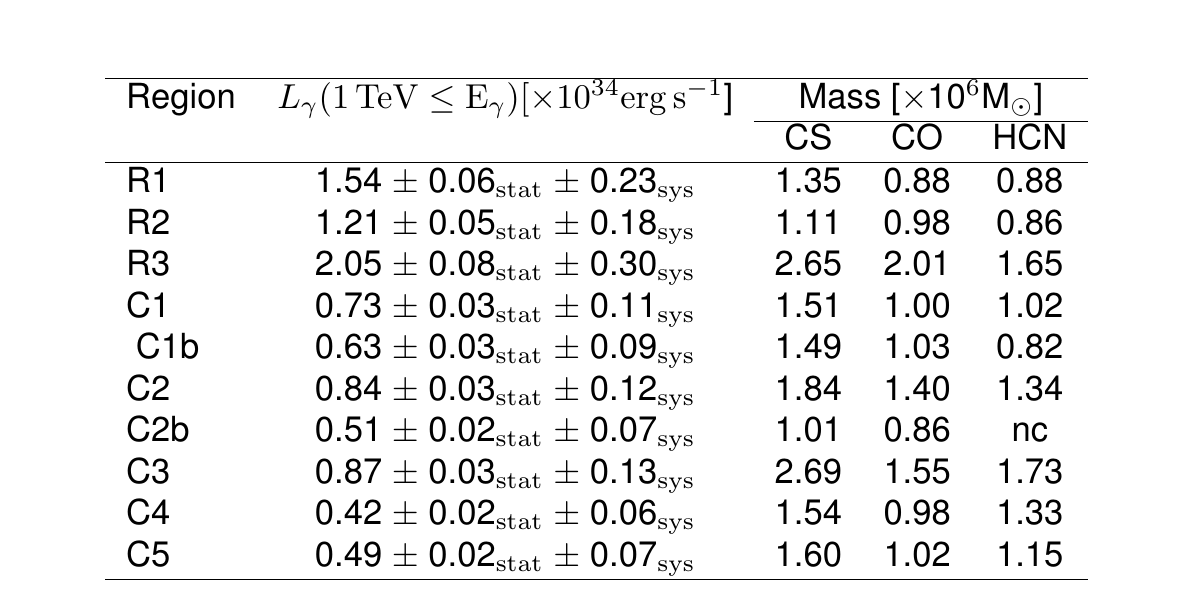} 
\caption{{\bf $\gamma$-ray luminosities and masses in different regions of the central molecular zone.} The errors quoted are 1$\sigma$ statistical (stat) and systematical (sys) errors.The region C2b does is not covered (nc) in the HCN line observations\cite{jones11}.}
\label{tab1}
\end{figure}
\vfill
\clearpage

 \begin{figure}[]
\includegraphics[width=1.0\textwidth]{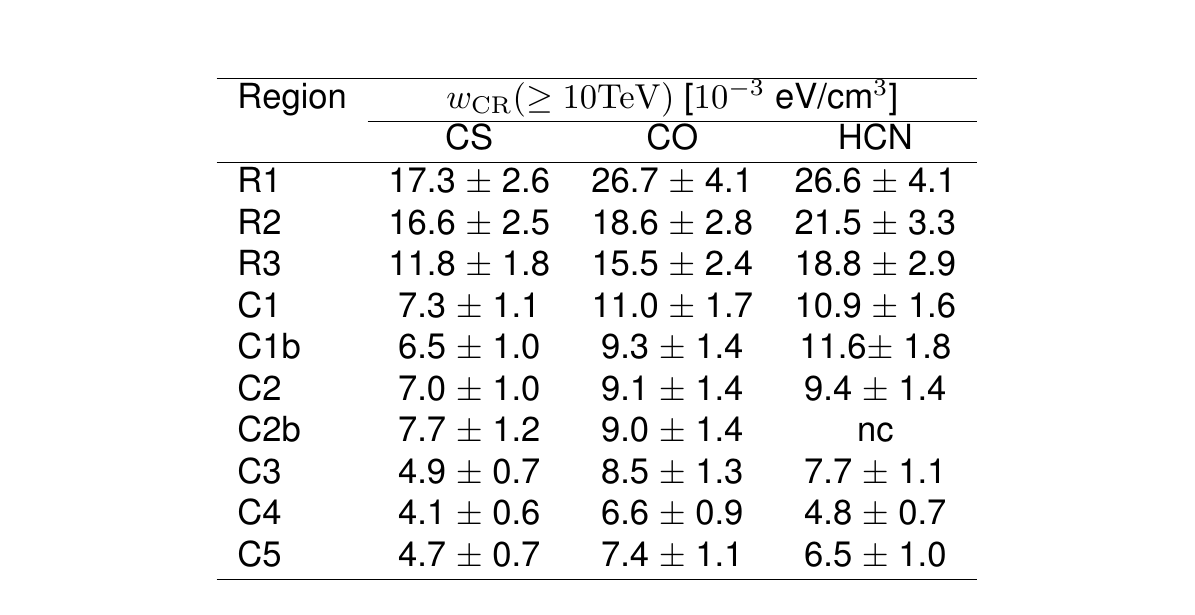} 
\caption{{\bf Cosmic-ray energy densities in different regions of the central molecular zone.} The densities are given in units of  $10^{-3}$ eV/cm$^{3}$, which is the value of the local cosmic-ray energy density measured in the Solar neighbourhood. The errors quoted are 1$\sigma$ statistical plus systematical errors. The region C2b does is not covered (nc) in the HCN line observations\cite{jones11}.}
\label{tab2}
\end{figure}
\vfill
\clearpage

 \begin{figure}[]
\includegraphics[width=1.0\textwidth]{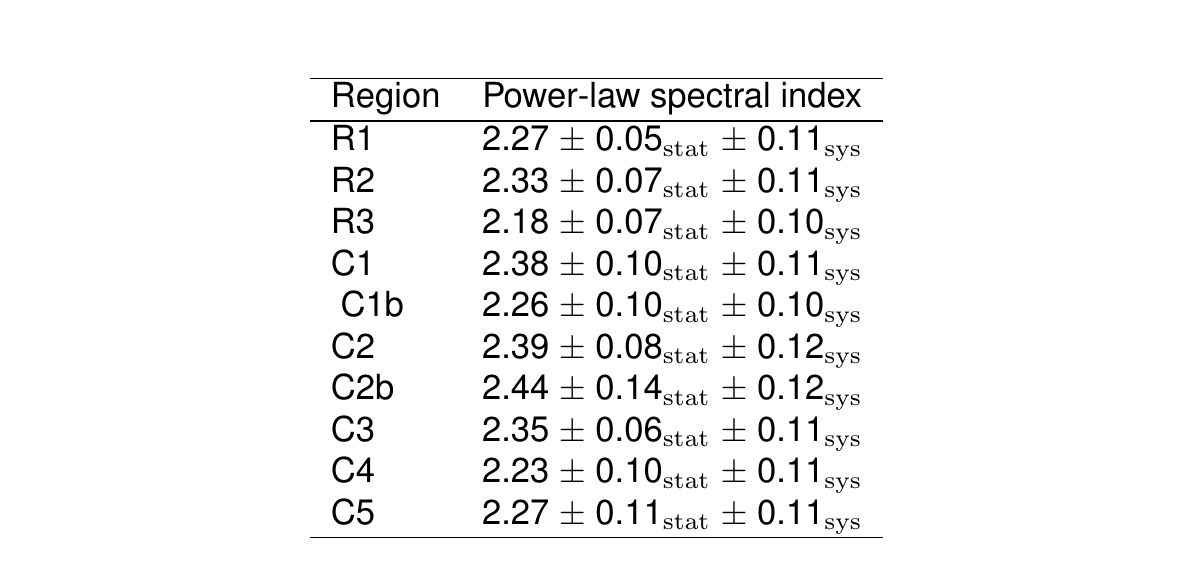} 
\caption{{\bf Power-law spectral indices of the $\gamma$-ray energy spectrum in different regions of the central molecular zone.} The quoted errors are 1$\sigma$ statistical (stat) and systematical (sys) errors.}
\label{tab3}
\end{figure}
\vfill
\clearpage

\renewcommand{\refname}{References}

\end{document}